\documentclass[aps,prd,twoside,twocolumn,nofootinbib,10pt,showpacs,floatfix]{revtex4-1}
\usepackage{amsmath,amssymb}
\usepackage{graphicx,bm}
\usepackage{slashed}
\usepackage{epstopdf}
\usepackage{ulem} 
\usepackage[usenames]{color}
\usepackage{float}
\usepackage{hyperref}
\usepackage{subfigure}
\usepackage{subfigure}
\usepackage{rotating}
\usepackage{color}
\usepackage{multirow}
\usepackage{dcolumn}
\usepackage{overpic}
\usepackage{booktabs}
\usepackage{makecell}
\usepackage{diagbox}

\renewcommand\sout{\bgroup \color{red} \ULdepth=-.5ex \ULset}

\newsavebox{\tablebox}
\begin{document}
\title{Discovery of $T_{c\bar s 0}^a(2900)^{0,++}$ implies new charmed-strange pentaquark system}
\author{Hong-Tao An$^{1,2}$}\email{anht14@lzu.edu.cn}
\author{Zhan-Wei Liu$^{1,2,3}$}\email{liuzhanwei@lzu.edu.cn}
\author{Fu-Sheng Yu$^{3,4,5}$}\email{yufsh@lzu.edu.cn}
\author{Xiang Liu$^{1,2,3}$}\email{xiangliu@lzu.edu.cn}
\affiliation{
$^1$School of Physical Science and Technology, Lanzhou University, Lanzhou 730000, China\\
$^2$Research Center for Hadron and CSR Physics, Lanzhou University and Institute of Modern Physics of CAS, Lanzhou 730000, China\\
$^3$Lanzhou Center for Theoretical Physics, and Frontiers Science Center for Rare Isotopes, Lanzhou University, Lanzhou 730000, China\\
$^4$School of Nuclear Science and Technology, Lanzhou University, Lanzhou 730000, China\\
$^5$Center for High Energy Physics, Peking University, Beijing 100871, China
}

\begin{abstract}
Inspired by the very recently discovered tetraquark states $T_{c\bar s 0}^a(2900)^{0,++}$ from the LHCb Collaboration, we predict the existence of a new charmed-strange pentaquark system, $c\bar s nnn$, which is closely connected to $c\bar s n\bar n$ by exchanging $\bar n$ into $nn$ with $n=u,d$. Especially, it is suggested to experimentally search for the predicted new pentaquark system via the weak decays of $B$ mesons or $b$-baryons, with the support from the study of 
the mass spectrum and the decay properties. 
The predicted new pentaquark system must attract extensive attention from experimentalists and theorists when it constructs the ``particle zoo 2.0" in the near future.
\end{abstract}
\maketitle

\section{Introduction}

The existence of multiquark{} states {was} proposed by Gell-Mann and Zweig \cite{GellMann:1964nj,Zweig:1981pd,Zweig:1964jf} at the birth of the quark model. The discoveries of these particles make the hadron world much {richer} in diversity. Their interesting properties will disclose the essence in the nonperturbative region of quantum chromodynamics (QCD) \cite{Chen:2016qju,Richard:2016eis,Hosaka:2016pey,Brambilla:2019esw,Chen:2022asf,Lebed:2016hpi}.

Among these reported new hadronic states in the past twenty years, the $D_{s0}(2317)^+$ cannot be passed
into silence, since its observation associated with
the $X(3872)$ in 2003 opened a new era of exploring exotic hadronic matter. The coupled-channel effect due to the $c\bar s n\bar n$ configuration was realized to solve the low-mass puzzle of $D_{s0}(2317)^+$ \cite{vanBeveren:2003kd,Dai:2003yg}. Here, for this excited charmed-strange meson, a pair of $n\bar n$ ($n=u, d$) is pulled out from the vacuum, which may renormalize the bare mass.
The isospin scalar $D_{s0}(2317)^+$ was observed in the $D_s^+\pi^0$ invariant spectrum in the channels of $B^{+}\to\bar D^{0}D_s^+\pi^{0}$ and $B^{0}\to\bar D^{-}D_s^+\pi^{0}$ \cite{BaBar:2003oey,Besson:2003cp}. The isospin vector mesons with charm and strangeness are explicitly exotic states and can couple to the $D_s\pi$ channel as well. 

When we are going to enter into the third decade since 2003, LHCb reported the observation of two isospin vector resonances, $T_{c\bar s 0}^a(2900)^{++}$ and $T_{c\bar s 0}^a(2900)^{0}$, in the $D_s^+\pi^\pm$ invariant spectrum of the similar channels $B^{+}\to\bar D^{-}D_s^+\pi^{+}$ and $B^{0}\to\bar D^{0}D_s^+\pi^{-}$ very recently \cite{TcsRept,LHCb:2022bkt}.  The least quark configurations of the $T_{c\bar s 0}^a(2900)^{++}$ and $T_{c\bar s 0}^a(2900)^{0}$ are obviously $c\bar s u \bar d$ and $c\bar s \bar u d$, respectively. There were some theoretical explorations of the $c\bar s n\bar n$ tetraquarks ever before  \cite{Chen:2017rhl,Agaev:2022duz,Lu:2020qmp,He:2020jna,Azizi:2018mte,Cheng:2020nho,Albuquerque:2020ugi,Guo:2021mja}. 

Standing at a new starting point, we should point out that the discoveries of the $T_{c\bar s 0}^a(2900)^{0,++}$ states help us study other similar multiquark systems more reliably, since it provides the critical understanding of the interactions among quarks. In some way, $nn$ is equivalent to $\bar n$ in the color space. Thus the $c\bar s nnn$ pentaquarks deserve a careful examination since it is closely related to $c\bar s n\bar n$ with the $\bar n$ replaced by $nn$. We will see in this work that the observation of $T_{c\bar s0}^a(2900)^{0,++}$ implies the existence of the charmed-strange pentaquarks, a new pentaquark system. The mass spectrum and decay properties of the $c\bar s nnn$ pentaquarks are systematically studied. Especially, we suggest several operational weak decay channels 
of $B$ mesons or $b$-baryons to search for the predicted charmed-strange pentaquark system, which is a crucial step of constructing the ``particle zoo 2.0".

\section{Mass spectrum of $c\bar{s}nnn$}

\begin{table*}[!]
\centering \caption{The masses and eigenvectors for the $c\bar{s}nnn$ pentaquark states. 
The explicit formulas of the basis vectors $|\Psi^{3/2}_{C1}\rangle$, $|\Psi^{3/2}_{D1}\rangle$... can be found in Appendix A of Ref. \cite{Weng:2019ynv}.
The masses are all in units of MeV.
}\label{mass}
\begin{lrbox}{\tablebox}
\renewcommand\arraystretch{1.05}
\renewcommand\tabcolsep{1pt}
\begin{tabular}{l|l|l|l}
\bottomrule[1.5pt]
\bottomrule[0.5pt]
$I(J^P)$&\multicolumn{1}{c|}{Eigenvector}&$I(J^P)$&\multicolumn{1}{c}{Eigenvector}\\
\bottomrule[0.7pt]
$\frac{3}{2}(\frac{3}{2}^{-})$&
$\begin{pmatrix}
|3530\rangle\\|3357\rangle\\|3199\rangle
\end{pmatrix}$=
$\begin{pmatrix}
0.792&-0.532&-0.301\\
0.397&0.822&-0.408\\
0.464&0.204&0.862
\end{pmatrix}$
$\begin{pmatrix}
|\Psi^{3/2}_{C1}\rangle\\ |\Psi^{3/2}_{C2}\rangle\\ |\Psi^{3/2}_{C3}\rangle
\end{pmatrix}$
&
$\frac{1}{2}(\frac{3}{2}^{-})$&
$\begin{pmatrix}
|3456\rangle\\|3404\rangle\\|3162\rangle\\|2954\rangle
\end{pmatrix}$=
$\begin{pmatrix}
-0.090&-0.741&-0.659&-0.095\\
0.741&-0.492&0.456&-0.026\\
0.398&0.195&-0.390&0.807\\
0.533&0.414&-0.454&-0.583
\end{pmatrix}$
$\begin{pmatrix}
|\Psi^{3/2}_{D1}\rangle\\ |\Psi^{3/2}_{D2}\rangle\\ |\Psi^{3/2}_{D3}\rangle \\ |\Psi^{3/2}_{D4}\rangle
\end{pmatrix}$
\\
$\frac{3}{2}(\frac{1}{2}^{-})$&
$\begin{pmatrix}
|3684\rangle\\|3546\rangle\\|3247\rangle
\end{pmatrix}$=
$\begin{pmatrix}
0.716&0.667&0.205\\
0.600&0.438&0.670\\
0.356&-0.603&0.714
\end{pmatrix}$
$\begin{pmatrix}
|\Psi^{1/2}_{C1}\rangle\\ |\Psi^{1/2}_{C2}\rangle\\ |\Psi^{1/2}_{C3}\rangle
\end{pmatrix}$
&
$\frac{1}{2}(\frac{1}{2}^{-})$&
$\begin{pmatrix}
|3425\rangle\\|3253\rangle\\|3116\rangle\\|2955\rangle\\|2828\rangle
\end{pmatrix}$=
$\begin{pmatrix}
0.529&-0.789&0.285&0.118&0.054\\
0.150&0.266&0.732&-0.575&-0.201\\
-0.530&-0.188&0.381&0.461&-0.571\\
0.360&0.485&0.347&0.664&0.273\\
-0.536&-0.193&0.342&-0.054&0.746
\end{pmatrix}$
$\begin{pmatrix}
|\Psi^{1/2}_{D1}\rangle\\ |\Psi^{1/2}_{D2}\rangle\\ |\Psi^{1/2}_{D3}\rangle \\ |\Psi^{1/2}_{D4}\rangle\\ |\Psi^{1/2}_{D5}\rangle
\end{pmatrix}$
\\
\bottomrule[0.5pt]
\bottomrule[1.5pt]
\end{tabular}
\end{lrbox}\scalebox{0.972}{\usebox{\tablebox}}
\end{table*}

The mass spectrum of the $c\bar{s}nnn$ pentaquark system is studied by considering the chromoelectric interactions (CEI) and the chromomagnetic interactions (CMI), which have been widely studied in various hadron systems \cite{An:2021vwi,An:2020vku,An:2020jix,Liu:2021gva,Weng:2021ngd,Guo:2021yws,Weng:2018mmf}. 
The effective Hamiltonian for the masses of $S$-wave hadrons can be described by 
\begin{eqnarray}\label{Eq3}
H&=&\sum_im_i+H_{\textrm{CEI}}+H_{\textrm{CMI}}\nonumber\\
&=&\sum_im_i-\sum_{i<j}A_{ij} \vec\lambda_i\cdot \vec\lambda_j-\sum_{i<j}v_{ij} \vec\lambda_i\cdot \vec\lambda_j \vec\sigma_i\cdot\vec\sigma_j \nonumber\\
&=&-\frac{3}{4}\sum_{i<j}m_{ij}V^{\rm C}_{ij}-\sum_{i<j}v_{ij}V^{\rm CMI}_{ij}.
\end{eqnarray}
Here, the subscripts $i,j$ represent the quarks or antiquarks.
$V^{\rm C}_{ij}=\vec\lambda_i\cdot \vec\lambda_j$ and $ V^{\rm CMI}_{ij}=\vec\lambda_i\cdot \vec\lambda_j \vec\sigma_i\cdot\vec\sigma_j$
are the color and chromomagnetic interactions between quarks, respectively, with $\sigma_{i}$ being the Pauli matrices and $\lambda_i$ the Gell-Mann matrices. 
$\lambda_i$s are replaced by $-\lambda_i^{*}$ for the antiquarks.
The mass parameters of quark pairs are
$m_{ij}=\frac{1}{4}(m_{i}+m_{j})+\frac{4}{3}A_{ij}$,
which contain the effective quark masses $m_{i,j}$ and the color interaction strengthes $A_{ij}$.
{The effective coupling constants between the $i$-th and $j$-th quarks, $v_{ij}$, determine the mass gaps.}
The values of $m_{ij}$ and $v_{ij}$ about the $c\bar{s}nnn$ system can be extracted from the conventional hadrons including $p$, $\Delta$, $\Lambda_c$, $\Sigma_c^{(*)}$, $D_s^{(*)}$, $K^{(*)}$, and the other $S$-wave baryons and mesons. The results are $m_{nn}=182.2$ MeV, $m_{n\bar s}=154.0$ MeV, $m_{nc}=520.0$ MeV, $m_{n\bar c}=519.0$ MeV, $v_{nn}=19.1$ MeV, $v_{n\bar s}=18.7$ MeV, $v_{nc}=3.9$ MeV, and $v_{c\bar s}=6.7$ MeV  \cite{An:2020vku,Weng:2021ngd,Weng:2018mmf}.  

The corresponding total wave functions have to be constructed in order to investigate the mass spectrum of the $c\bar{s}nnn$ pentaquark states.
The total wave functions include the spatial part, the flavor part, the spin part and the color part, i.e. $\psi_{\rm tot}=\psi_{\rm space}\otimes\psi_{\rm flavor}\otimes\psi_{\rm spin}\otimes\psi_{\rm color}$.
They should be fully antisymmetric when exchanging identical quarks.
{The explicit expressions of the total wave functions can be obtained by the similar way of deducing the wave functions of the $c\bar{c}nnn$ pentaquark system (see Appendix A of Ref. \cite{Weng:2019ynv} for more details).
By considering the similarity of structures for the $c\bar{c}nnn$ and $c\bar{s}nnn$ states, the masses of $c\bar{s}nnn$ are also be obtained from the eigenvalue equations, where the quark flavor of $\bar{c}$ should be replaced by $\bar{s}$. 
Thus the different masses with the same $I(J^P)$ could be distinguished by the corresponding eigenvectors.}
Using the above values of parameters extracted by the conventional hadrons, the mass of a $J^{P}=0^{+}$ isovector $c\bar{s}n\bar{n}$ tetraquark states is given as 2863 MeV, which is very close to the measured value of $M_{T^{a}_{c\bar{s}0}(2900)}=2908\pm11\pm20$ MeV \cite{TcsRept}. 
{It manifests that the CEI and CMI model is suitable to describe the charmed-strange tetraquark. }
The central value of the mass of $T^{a}_{c\bar{s}0}(2900)$ can be obtained by overall increasing the parameters $m_{ij}$ by 1.5$\%$.

Taking $T^{a}_{c\bar{s}0}(2900)$ as the benchmark, we predict the mass spectrum of the $c\bar{s}nnn$ pentaquark states of $I(J^P)=1/2(1/2^-)$, $1/2(3/2^-)$, $3/2(1/2^-)$ and $3/2(3/2^-)$. {The masses and eigenvectors of the  $c\bar{s}nnn$ pentaquark states are shown in Table \ref{mass}.  The explicit formulas of the basis vectors $|\Psi^{3/2}_{C1}\rangle$, $|\Psi^{3/2}_{D1}\rangle$... can be found in the Appendix A of Ref. \cite{Weng:2019ynv}.} The mass region is between 2828 MeV and 3684 MeV, which is accessible in the weak decays of $b$-hadrons.
The $c\bar{s}nnn$ states can be easily recognized as an explicitly exotic candidate once observed, since they have no quark-antiquark pair with the same flavor and thus are impossible to be mixed with the ordinary baryons.

\begin{table*}[t]
\centering \caption{The decay properties of the $c\bar snnn$ pentaquark states. 
It is given for each possible decay channel by the values of overlaps $c_i$ and  $k\cdot|c_{i}^{2}|$, whose physical meanings are given in the main text. The overlaps are dimensionless, while the masses and $k\cdot|c_{i}^{2}|$'s are in the unit of MeV.
The empty entries denote that the focused pentaquark state cannot decay into the final state by the S-wave rearrangment.
The kinetically forbidden decay channels are marked with ``$\times$”. 
According to the values of $k\cdot|c_{i}^{2}|$, one can roughly estimate the relative decay widths between different decay processes of individual initial pentaquark states in the each block.
}\label{csnnn}
\begin{lrbox}{\tablebox}
\renewcommand\arraystretch{1.06}
\renewcommand\tabcolsep{2.25pt}
\begin{tabular}{cl|cc|cccccc|cc|cccccc}
\bottomrule[1.5pt]
\bottomrule[0.5pt]
\multicolumn{2}{c|}{$c\bar{s}nnn$}&\multicolumn{8}{c|}{Overlaps $c_i$}&\multicolumn{7}{c}{Values of $k\cdot|c_{i}^{2}|$}&\\
\multicolumn{2}{l|}{}&\multicolumn{2}{c|}{$nnn\bigotimes c\bar{s}$}&\multicolumn{6}{c|}{$nnc\bigotimes n\bar{s}$}&\multicolumn{2}{c|}{$nnn\bigotimes c\bar{s}$}&\multicolumn{6}{c}{$nnc\bigotimes n\bar{s}$}\\
$I(J^P)$&Mass&$\Delta D_{s}^{*}$&$\Delta D_{s}$&
$\Sigma^{*}_{c}K^{*}$&$\Sigma^{*}_{c}K$&
$\Sigma_{c}K^{*}$&$\Sigma_{c}K$&
$\Lambda_{c}K^{*}$&$\Lambda_{c}K$
&$\Delta D_{s}^{*}$&$\Delta D_{s}$&
$\Sigma^{*}_{c}K^{*}$&$\Sigma_{c}K^{*}$&
$\Sigma^{*}_{c}K$&$\Sigma_{c}K$&
$\Lambda_{c}K^{*}$&$\Lambda_{c}K$
\\
\bottomrule[0.7pt]
$\frac{3}{2}(\frac{3}{2}^{-})$
&3530&-0.532&-0.301&0.488&-0.140&-0.250&&&&155&66&97&32&\multicolumn{1}{|c}{17}&&\multicolumn{1}{|c}{}&\multicolumn{1}{|c}{}\\
&3357&0.822&-0.408&-0.239&0.262&-0.195&&&&96&82&$\times$&5&\multicolumn{1}{|c}{41}&&\multicolumn{1}{|c}{}&\multicolumn{1}{|c}{}\\
&3199&0.204&0.862&-0.034&0.366&0.218&&&&$\times$&$\times$&$\times$&$\times$&\multicolumn{1}{|c}{56}&&\multicolumn{1}{|c}{}&\multicolumn{1}{|c}{}\\
$\frac{3}{2}(\frac{1}{2}^{-})$
&3684&0.205&&0.611&&-0.238&-0.028&&&32&&234&39&\multicolumn{1}{|c}{}&\multicolumn{1}{c|}{1}&&\multicolumn{1}{|c}{}\\
&3546&0.670&&0.124&&0.523&-0.079&&&256&&7&144&\multicolumn{1}{|c}{}&\multicolumn{1}{c|}{5}&&\multicolumn{1}{|c}{}\\
&3247&0.714&&0.137&&0.202&-0.464&&&$\times$&&$\times$&$\times$&\multicolumn{1}{|c}{}&\multicolumn{1}{c|}{117}&&\multicolumn{1}{|c}{}\\
\Xcline{3-4}{0.7pt}\Xcline{11-12}{0.7pt}
&&$N D_{s}^{*}$&$N D_{s}$&&&&&&&$N D_{s}^{*}$&$N D_{s}$&&&\multicolumn{1}{|c}{}&\multicolumn{1}{c|}{}&&\multicolumn{1}{|c}{}\\
\Xcline{3-4}{0.7pt}\Xcline{11-12}{0.7pt}
$\frac{1}{2}(\frac{3}{2}^{-})$
&3456&-0.095&&-0.816&-0.132&-0.183&&\multicolumn{1}{c}{$<10^{-4}$}&&7&&166&13&\multicolumn{1}{|c}{12}&\multicolumn{1}{c|}{}&$\times$&\multicolumn{1}{|c}{}\\
&3403&-0.026&&0.037&-0.049&0.887&&\multicolumn{1}{c}{$<10^{-4}$}&&0.5&&$\times$&220&\multicolumn{1}{|c}{2}&\multicolumn{1}{c|}{}&$\times$&\multicolumn{1}{|c}{}\\
&3162&0.807&&-0.188&-0.329&0.177&&\multicolumn{1}{c}{$<10^{-4}$}&&251&&$\times$&$\times$&\multicolumn{1}{|c}{40}&\multicolumn{1}{c|}{}&$\times$&\multicolumn{1}{|c}{}\\
&2954&-0.583&&-0.035&0.806&0.016&&\multicolumn{1}{c}{$<10^{-4}$}&&$\times$&&$\times$&$\times$&\multicolumn{1}{|c}{$\times$}&\multicolumn{1}{c|}{}&$\times$&\multicolumn{1}{|c}{}\\
$\frac{1}{2}(\frac{1}{2}^{-})$
&3425&0.118&0.054&0.539&&0.518&0.092&0.555&0.055&10&2&43&88&\multicolumn{1}{|c}{}&\multicolumn{1}{c|}{6}&180&\multicolumn{1}{|c}{3}\\
&3253&-0.575&-0.201&0.472&&-0.563&-0.041&0.053&-0.310&174&28&$\times$&$\times$&\multicolumn{1}{|c}{}&\multicolumn{1}{c|}{1}&1&\multicolumn{1}{|c}{68}\\
&3116&0.461&-0.571&-0.267&&-0.171&-0.404&-0.477&0.196&63&173&$\times$&$\times$&\multicolumn{1}{|c}{}&\multicolumn{1}{c|}{65}&$\times$&\multicolumn{1}{|c}{22}\\
&2955&0.664&0.273&0.081&&-0.101&-0.391&0.133&0.600&$\times$&19&$\times$&$\times$&\multicolumn{1}{|c}{}&\multicolumn{1}{c|}{13}&$\times$&\multicolumn{1}{|c}{143}\\
&2828&-0.054&0.745&-0.024&&-0.076&0.672&0.002&-0.242&$\times$&$\times$&$\times$&$\times$&\multicolumn{1}{|c}{}&\multicolumn{1}{c|}{$\times$}&$\times$&\multicolumn{1}{|c}{12}\\
\bottomrule[0.5pt]
\bottomrule[1.5pt]
\end{tabular}
\end{lrbox}\scalebox{1.015}{\usebox{\tablebox}}
\end{table*}

\begin{figure*}[t]
\centering
\includegraphics[width=17.5cm]{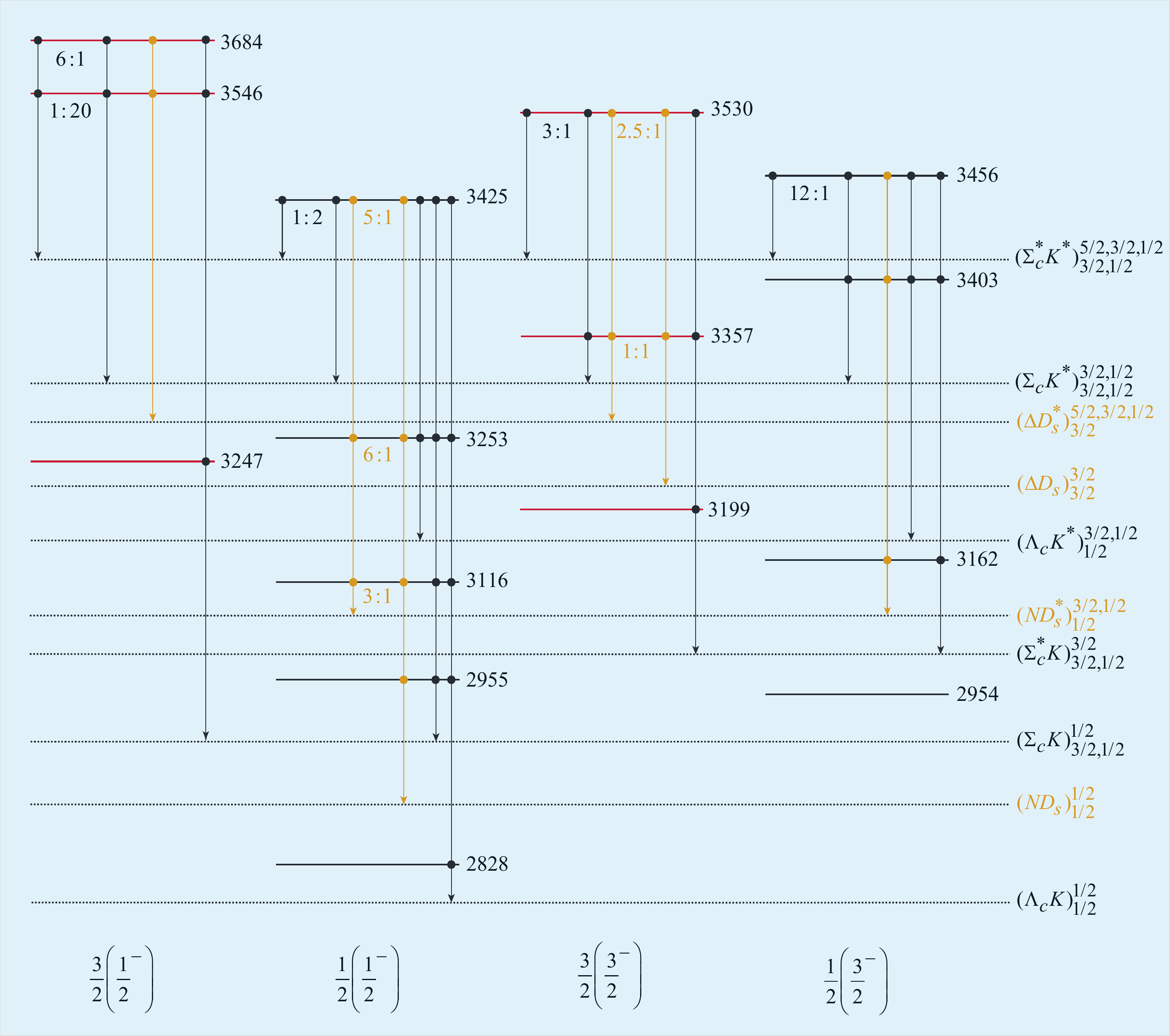}\\
\caption{Schematic diagram of the mass spectrum and the decay properties of the  $c\bar{s}nnn$ pentaquark states.
The relative mass positions (units: MeV) of the pentaquark states with $I=3/2$ and $I=1/2$  are labeled by the red and black horizontal solid lines, respectively.
The dotted lines denote the various baryon-meson thresholds. The superscripts (subscripts) of the thresholds, e.g. $(\Sigma^{*}_{c}K^{*})^{5/2,3/2,1/2}_{3/2,1/2}$, represent the possible total angular momenta (isospins) of the relevant states.
The vertical solid lines with arrows indicate the kinematically allowed decaying processes with the rearranged $S$-wave interactions. The orange and the black vertical lines correspond to the $nnn$-$c\bar{s}$ and $nnc$-$n\bar{s}$ modes, respectively. The ratios between two vertical solid lines with arrows represent the relative partial widths from the left to the right decaying processes.
}\label{figcsnnn}
\end{figure*}

\section{Decays of $c\bar s nnn$}

The decay properties are another important aspect to investigate the nature of exotic hadrons. 
The decays of the $c\bar s nnn$ pentaquarks can be studied by the overlaps using the above effective Hamiltonian and wave functions. The overlap between a pentaquark state and a specific baryon $\otimes$ meson state is defined by transforming the eigenvectors of the pentaquark states into the baryon-meson configurations, such as $c_i=\langle N\otimes D_s|P_{c\bar sn^3}(2828)\rangle$.
The baryon and meson components inside the pentaquarks can be divided into the $1\otimes 1$ components and the $8 \otimes 8$  components in the color space. 
The $1\otimes 1$ component can be easily dissociated into an $S$-wave baryon and an $S$-wave meson, which are
OZI-superallowed decay processes. 
On the contrary, the $8 \otimes 8$ component cannot fall apart if without the gluon exchange force.
Thus, we will only concentrate on the  OZI-superallowed strong decay processes. The values of overlaps are shown in Table. \ref{csnnn}.

The partial width of a two-body $L$-wave decay reads as \cite{Weng:2019ynv,Weng:2020jao,Weng:2021hje}
\begin{eqnarray}\label{width}
\Gamma_{i}=\gamma_{i}\alpha\frac{k^{2L+1}}{m^{2L}}\,|c_{i}|^{2},
\end{eqnarray}
where $\alpha$ is an effective coupling constant, $c_{i}$ is the overlap,  $m$ is the mass of initial state, $k$ is the momentum of final state in the rest frame of the initial state, and $\gamma_i$ is a factor depending on the decay processes to be discussed later. 
Since $(k/m)^{2}$ is of the order of $10^{-2}$ or even smaller, the contributions from higher partial-wave decays would be suppressed. 
Thus, we only consider the $S$-wave strong decays, whose widths are simplified as $\Gamma_{i}=\gamma_{i}\alpha k|c_{i}|^{2}$.
The values of $k\cdot|c_{i}|^{2}$ for each decay channel are shown in Table \ref{csnnn}.

According to the obtained $c\bar{s}nnn$ mass spectrum, we plot their relative positions associated with all possible rearranged decay channels in Fig. \ref{figcsnnn}.
It can be seen that the $c\bar{s}nnn$ system has 10 possible two-body strong decay channels of $\Sigma_c^{(*)}K^{(*)}$, $\Lambda_cK^{(*)}$, $\Delta D_s^{(*)}$ and $ND_s^{(*)}$.
The angular momenta (isospins) of the baryon-meson states are labeled by superscripts (subscripts) in the rearranged decay channels. 
If both the spin and the isospin of an initial pentaquark state are equal to the numbers in the superscript and subscript of a baryon-meson state respectively, 
this pentaquark could decay into the corresponding baryon-meson channel via the $S$-wave strong interaction. 
{The vertical solid lines with arrows indicate the kinematically allowed decaying processes with the rearranged $S$-wave interactions. The orange and the black vertical lines correspond to the $nnn$-$c\bar{s}$ and $nnc$-$n\bar{s}$ modes, respectively.
}

All of the $I(J^{P})=1/2(1/2^-)$ pentaquark states can decay into $\Lambda_cK$. Except for the lowest lying state $\rm P_{c\bar{s}n^{3}}(2828)$, all the other $1/2(1/2^-)$ states can also decay into $ND_s$ and $\Sigma_cK$. Similarly, all the $3/2(1/2^-)$ states can decay into $\Sigma_cK$, while all the $1/2(3/2^-)$ and $3/2(3/2^-)$ states except for $\rm P_{c\bar{s}n^{3}}(2954)$ can decay into $\Sigma_c^*K$. Note that both the $\Sigma_c$ and $\Sigma_c^*$ totally decay into $\Lambda_c\pi$ which are easily reconstructed in experiments. Therefore, most of the $c\bar s nnn$ pentaquark states can be measured via either $\Lambda_cK$, $ND_s$, $\Sigma_cK$ or $\Sigma_c^*K$. These decaying properties might be helpful for experimental searches. 

{The $\rm P_{c\bar{s}n^{3}}(2828)$ state can only decay into $\Lambda_{c}K$ with a small phase space, leading to be a relative narrow state.
Below all the allowed rearranged $S$-wave decay channels, the $\rm P_{c\bar{s}n^{3}}(2954)$ state can decay into $ND_{s}$, $\Lambda_cK$ or $\Sigma_cK$ only via $D$-wave interactions whose widths are not available in this work. Thus the decay channels of $\rm P_{c\bar{s}n^{3}}(2954)$ are not shown in Fig. \ref{figcsnnn}. 
}

Most of the other pentaquark states have more than two decay channels. 
With the values of $k\cdot|c_i|^2$ in Table. \ref{csnnn}, we can compare the partial decay widths if the $\gamma_i$s are known. 
{The factor of $\gamma_{i}$ is a process-related quantity depending on the spatial wave functions of the initial pentaquark states and the final-state mesons and baryons.}
The spatial wave functions of the pseudoscalar meson $D_{s}$ and the vector meson $D^{*}_{s}$ are nearly the same in the quark model.
The deviations between the spatial wave functions of $\Sigma_{c}$ and $\Sigma^{*}_{c}$ could be neglected in the heavy quark limit \cite{Weng:2019ynv}.
The relationships of $\gamma_{i}$ in some decay channels can thus be taken approximately as
$\gamma_{\Delta D^{*}_{s}}=\gamma_{\Delta D_{s}}$, 
$\gamma_{N D^{*}_{s}}=\gamma_{N D_{s}}$, $\gamma_{\Sigma_{c} K^{*}}=\gamma_{\Sigma^{*}_{c} K^{*}}$, and
$\gamma_{\Sigma_{c}K} =\gamma_{\Sigma^{*}_{c} K}$.
{Then, one can roughly estimate the ratios of the partial widths for an individual pentaquark state decaying into a pair of final states shown in one block of $k\cdot|c_i|^2$ in Table \ref{csnnn}.
For example, the ratio of the partial widths,  
$\Gamma_{P_{c\bar{s}n^{3}}(3357)\to\Delta D_{s}}:\Gamma_{P_{c\bar{s}n^{3}}(3357)\to \Delta D^{*}_{s}}$, is approximate to be $1:1$, given by the values of $k\cdot|c_i|^2$ as 96 and 82, respectively.}
 Such ratios of the relative branching fractions are provided between the vertical lines of the relevant processes in Fig. \ref{figcsnnn}. 
These ratios can help to guide for the experimental searches, and to investigate the inner structures of the pentaquark states by the future measurements.



\section{Suggestions on the experimental searches}

\begin{figure}[!]
\centering
\begin{tabular}{c}
\includegraphics[width=7cm]{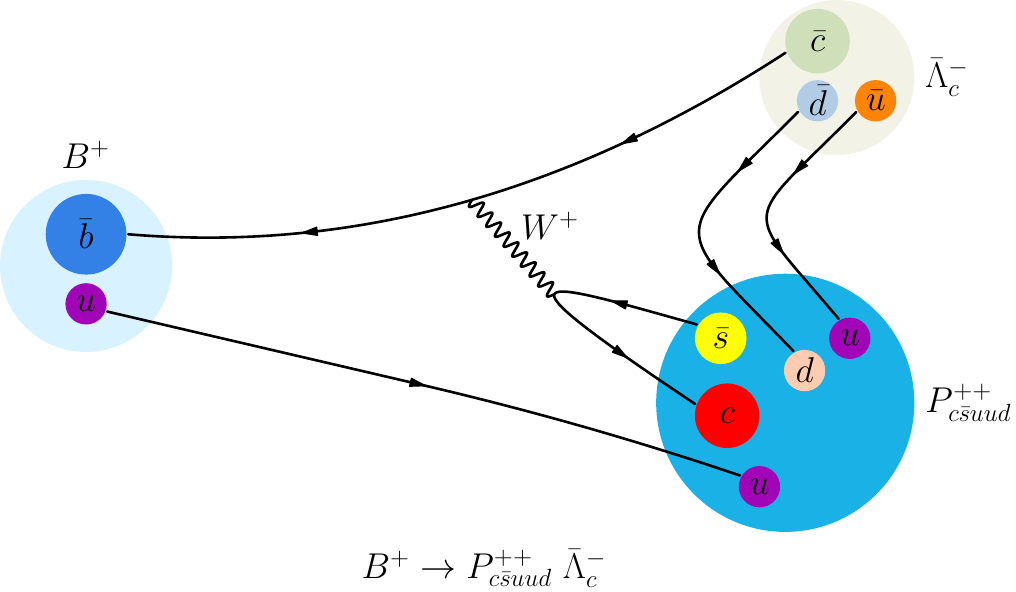}\\
\includegraphics[width=7cm]{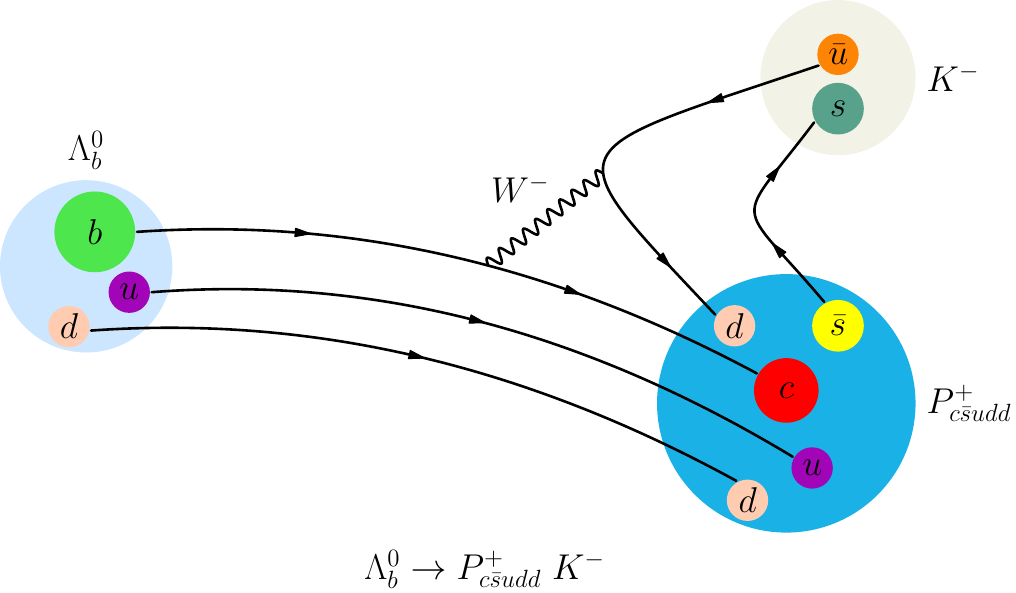}\\
\end{tabular}
\caption{
The production processes $B^+\to P_{c\bar{s}uud}^{++}\bar{\Lambda}_c^-$ (top)
and $\Lambda_b^0\to P_{c\bar{s}udd}^+K^-$ (bottom) at the quark level.
}\label{decay}
\end{figure}

In analogy to $T_{c\bar s0}(2900)^{0,++}$, the charmed-strange pentaquark states can also be observed in the decays of $b$ hadrons. The largest production processes are among the transitions of $b\to c\bar c s$ and $b\to c \bar u d$ which are the most Cabibbo-Kobayashi-Maskawa(CKM)-favored, with $|V_{cb}V_{cs}^*|\approx|V_{cb}V_{ud}^*|\approx\lambda^2$ and the Wolfenstein parameter $\lambda\approx0.225$. We propose to search for the charmed-strange pentaquark states via $B^+\to P_{c\bar suud}^{++}\bar \Lambda_c^-$ and $\Lambda_b^0\to P_{c\bar sudd}^{+} K^-$, respectively. The decaying topological diagrams of the above processes are given in Fig. \ref{decay}.
The branching fraction of $\Lambda_b^0\to P_{c\bar sudd}^{+} K^-$ is expected to be at the order of $10^{-5}$, similar to the ones of $\Lambda_b^0\to P_\psi^N(4312)^+K^-$ and  $\Lambda_b^0\to P_\psi^N(4440)^+K^-$ with the same topological diagrams \cite{Chen:2020eyu} by just exchanging from the $b\to c\bar cs$ transition into the $b\to c\bar u d$ and from the $u\bar u$ pair from vacuum into the $s\bar s$. The branching fraction of $B^+\to P_{c\bar suud}^{++}\bar \Lambda_c^-$ is about 1 order of magnitude smaller due to the one more quark-antiquark pair from the vacuum, thus at the order of $\mathcal{O}(10^{-6})$.
Some other processes are also suggested, such as $B^+\to P_{c\bar suud}^{++}\bar p$ and $\bar B_s^0\to P_{c\bar sudd}^{+}\bar p$.
Considering the experimental conditions with charged particles preferred, the charmed-strange pentaquark states can be reconstructed via $P_{c\bar suud}^{++}\to D_s^+p$ or $\Lambda_c^+K^+$, and $P_{c\bar sudd}^{+}\to \Sigma_c^0K^+$ or $\Sigma_c^{*0}K^+$ and subsequently into $(\Lambda_c^+\pi^-)K^+$. 
Note that the above $P_{c\bar suud}^{++}$ and $P_{c\bar sudd}^{+}$ represent not only any special pentaquark state, but all the possible states with the corresponding quark flavors. 
Considering the large data of $b$ hadrons at LHCb and the observations of $T_{c\bar s0}(2900)^{0,++}$ and some other tetraquark and pentaquark candidates in the decays of $B$ mesons or $b$-baryons, 
{we hope to observe} the charmed-strange pentaquarks via the mentioned processes.

\section{Summary}

{After the first observation of the charmed-strange meson $D_{s0}(2317)$ for around twenty years,}
the two new observed charmed-strange tetraquark states $T_{c\bar s 0}^a(2900)^{0}$ and $T_{c\bar s 0}^a(2900)^{++}$ are constructing the ``particle zoo 2.0", together with many other recent observed exotic states. 
In this work, we point out that the discovery of $T_{c\bar s 0}^a(2900)^{0,++}$ implies the existence of their closely related charmed-strange pentaquarks $c\bar snnn$, a new  pentaquark system.
We have systematically studied the mass spectrum and the decay properties of the $c\bar{s}nnn$ states. 
Some suggestions are given for the experimental searches in the $b$-hadron decays, such as $B^+\to P_{c\bar suud}^{++}\bar \Lambda_c^-$ and $\Lambda_b^0\to P_{c\bar sudd}^{+} K^-$ with the pentaquarks reconstructed by $P_{c\bar suud}^{++}\to D_s^+p$ or $\Lambda_c^+K^+$, and $P_{c\bar sudd}^{+}\to \Sigma_c^0K^+$ or $\Sigma_c^{*0}K^+$.
The predicted new pentaquark system must attract extensive attention from experimentalists and theorists in the near future.

\vfil
\section*{ACKNOWLEDGMENTS}
This work is supported by the China National Funds for Distinguished Young Scientists under Grant No. 11825503, National Key Research and Development Program of China under Contract No. 2020YFA0406400, the 111 Project under Grant No. B20063, CAS Interdisciplinary Innovation Team, and the National Natural Science Foundation of China under Grant Nos. 12175091, 11975112, 11965016, and 12047501.


\begin{thebibliography}{300}
\bibitem{GellMann:1964nj}
M.~Gell-Mann,
Phys. Lett. \textbf{8}, 214-215 (1964).

\bibitem{Zweig:1981pd}
G.~Zweig,
Version 1, CERN-TH-401.

\bibitem{Zweig:1964jf}
G.~Zweig,
Version 2, CERN-TH-412.
  
\bibitem{Hosaka:2016pey}
A.~Hosaka, T.~Iijima, K.~Miyabayashi, Y.~Sakai and S.~Yasui,
PTEP \textbf{2016}, 062C01 (2016),
[arXiv:1603.09229 [hep-ph]].
\bibitem{Chen:2016qju}
H.~X.~Chen, W.~Chen, X.~Liu and S.~L.~Zhu,
Phys. Rept. \textbf{639}, 1-121 (2016),
[arXiv:1601.02092 [hep-ph]].
\bibitem{Richard:2016eis}
J.~M.~Richard,
Few Body Syst. \textbf{57}, 1185-1212 (2016),
[arXiv:1606.08593 [hep-ph]].

\bibitem{Lebed:2016hpi}
R.~F.~Lebed, R.~E.~Mitchell and E.~S.~Swanson,
Prog. Part. Nucl. Phys. \textbf{93} (2017), 143-194,
[arXiv:1610.04528 [hep-ph]].

\bibitem{Brambilla:2019esw}
N.~Brambilla, S.~Eidelman, C.~Hanhart, A.~Nefediev, C.~P.~Shen, C.~E.~Thomas, A.~Vairo and C.~Z.~Yuan,
Phys. Rept. \textbf{873} (2020), 1-154,
[arXiv:1907.07583 [hep-ex]].

\bibitem{Chen:2022asf}
H.~X.~Chen, W.~Chen, X.~Liu, Y.~R.~Liu and S.~L.~Zhu,
[arXiv:2204.02649 [hep-ex]].

\bibitem{vanBeveren:2003kd}
E.~van Beveren and G.~Rupp,
Phys. Rev. Lett. \textbf{91}, 012003 (2003).



\bibitem{Dai:2003yg}
Y.~B.~Dai, C.~S.~Huang, C.~Liu and S.~L.~Zhu,
Phys. Rev. D \textbf{68}, 114011 (2003).

\bibitem{BaBar:2003oey}
B.~Aubert \textit{et al.} [BaBar],
Phys. Rev. Lett. \textbf{90} (2003), 242001,
[arXiv:hep-ex/0304021 [hep-ex]].

\bibitem{Besson:2003cp}
  D.~Besson {\it et al.} [CLEO Collaboration],
  Phys.\ Rev.\ D {\bf 68}, 032002 (2003),
  [hep-ex/0305100].

\bibitem{TcsRept}
 [LHCb],
[arXiv:2212.02716 [hep-ex]].

\bibitem{LHCb:2022bkt}
 [LHCb],
[arXiv:2212.02717 [hep-ex]].


\bibitem{Agaev:2022duz}
S.~S.~Agaev, K.~Azizi, and H.~Sundu,
arXiv:2207.02648 [hep-ex].

\bibitem{Azizi:2018mte}
K.~Azizi and U.~\"Ozdem,
J. Phys. G \textbf{45} (2018) no.5, 055003,
[arXiv:1802.07711 [hep-ph]].

\bibitem{Lu:2020qmp}
Q.~F.~L\"u, D.~Y.~Chen and Y.~B.~Dong,
Phys. Rev. D \textbf{102} (2020) no.7, 074021,
[arXiv:2008.07340 [hep-ph]].

\bibitem{He:2020jna}
X.~G.~He, W.~Wang and R.~Zhu,
Eur. Phys. J. C \textbf{80}, no.11, 1026 (2020),
[arXiv:2008.07145 [hep-ph]].

\bibitem{Cheng:2020nho}
J.~B.~Cheng, S.~Y.~Li, Y.~R.~Liu, Y.~N.~Liu, Z.~G.~Si and T.~Yao,
Phys. Rev. D \textbf{101} (2020) no.11, 114017,
[arXiv:2001.05287 [hep-ph]].

\bibitem{Albuquerque:2020ugi}
R.~M.~Albuquerque, S.~Narison, D.~Rabetiarivony and G.~Randriamanatrika,
Nucl. Phys. A \textbf{1007} (2021), 122113,
[arXiv:2008.13463 [hep-ph]].

\bibitem{Guo:2021mja}
T.~Guo, J.~Li, J.~Zhao and L.~He,
Phys. Rev. D \textbf{105}, no.5, 054018 (2022),
[arXiv:2108.06222 [hep-ph]].


\bibitem{Chen:2017rhl}
W.~Chen, H.~X.~Chen, X.~Liu, T.~G.~Steele and S.~L.~Zhu,
Phys. Rev. D \textbf{95} (2017) no.11, 114005,
[arXiv:1705.10088 [hep-ph]].

\bibitem{An:2020jix}
H.~T.~An, K.~Chen, Z.~W.~Liu and X.~Liu,
Phys. Rev. D \textbf{103} (2021) no.7, 074006,
[arXiv:2012.12459 [hep-ph]].

\bibitem{An:2021vwi}
H.~T.~An, K.~Chen, Z.~W.~Liu and X.~Liu,
Phys. Rev. D \textbf{103} (2021) no.11, 114027,
[arXiv:2106.02837 [hep-ph]].

\bibitem{Liu:2021gva}
Z.~Liu, H.~T.~An, Z.~W.~Liu and X.~Liu,
Phys. Rev. D \textbf{105} (2022) no.3, 034006,
[arXiv:2112.02510 [hep-ph]].

\bibitem{Guo:2021yws}
T.~Guo, J.~Li, J.~Zhao and L.~He,
Phys. Rev. D \textbf{105} (2022) no.1, 014021,
[arXiv:2108.10462 [hep-ph]].

\bibitem{Weng:2018mmf}
X.~Z.~Weng, X.~L.~Chen and W.~Z.~Deng,
Phys. Rev. D \textbf{97} (2018) no.5, 054008,
[arXiv:1801.08644 [hep-ph]].


\bibitem{Weng:2021ngd}
X.~Z.~Weng, W.~Z.~Deng and S.~L.~Zhu,
Phys. Rev. D \textbf{105} (2022) no.3, 034026,
[arXiv:2109.05243 [hep-ph]].


\bibitem{An:2020vku}
H.~T.~An, K.~Chen and X.~Liu,
Phys. Rev. D \textbf{105} (2022) no.3, 034018,
[arXiv:2010.05014 [hep-ph]].

\bibitem{Weng:2019ynv}
X.~Z.~Weng, X.~L.~Chen, W.~Z.~Deng and S.~L.~Zhu,
Phys. Rev. D \textbf{100} (2019) no.1, 016014,
[arXiv:1904.09891 [hep-ph]].







\bibitem{Weng:2020jao}
X.~Z.~Weng, X.~L.~Chen, W.~Z.~Deng and S.~L.~Zhu,
Phys. Rev. D \textbf{103}, no.3, 034001 (2021),
[arXiv:2010.05163 [hep-ph]].

\bibitem{Weng:2021hje}
X.~Z.~Weng, W.~Z.~Deng and S.~L.~Zhu,
Chin. Phys. C \textbf{46}, no.1, 013102 (2022),
[arXiv:2108.07242 [hep-ph]].


\bibitem{Chen:2020eyu}
Y.~K.~Chen, J.~J.~Han, Q.~F.~L\"u, J.~P.~Wang and F.~S.~Yu,
Eur. Phys. J. C \textbf{81} (2021) no.1, 71,
[arXiv:2009.01182 [hep-ph]].







\end{thebibliography}
\end{document}